# Polariton Emission Characteristics of a Modulation-Doped Multiquantum-Well Microcavity Diode

Ayan Das, Bo Xiao, Sishir Bhowmick and Pallab Bhattacharya*

*Center for Photonics and Multiscale Nanomaterials, Department of Electrical Engineering and Computer Science, University of Michigan, Ann Arbor, Michigan 48109-2122, USA*

The role of polariton-electron scattering on the performance characteristics of an electrically injected GaAs-based quantum well microcavity diode in the strong coupling regime has been investigated. An electron gas is introduced in the quantum wells by modulation doping with silicon dopants. It is observed that polariton-electron scattering suppresses the relaxation bottleneck in the lower polariton branch. However, it is not adequate to produce a degenerate coherent condensate at $k_{||} \sim 0$ and coherent emission.

* pkb@eecs.umich.edu

Strong coupling between quantum well (QW) excitons and photons confined in a microcavity leads to the creation of polaritons which are bosons [1-3]. Although polariton condensation [4-6] and room temperature coherent polariton emission [7-10] have been demonstrated in various material systems, the excitation of QW excitons was achieved via optical pumping. For device applications, it is necessary to demonstrate polariton condensation and coherent emission with non-resonant electrical injection, which still remains a challenge. Over the past few years, polariton emission in electrically injected *p-i-n* microcavities has been reported for GaAs structures [11-13], and electrically injected polariton LEDs operating at room temperature have also been demonstrated recently [14].

The difficulty in achieving a polariton condensate with non-resonant excitation lies in overcoming the relaxation bottleneck, which is the result of a decrease in the density of states of the lower polaritons (LPs) when their characteristics on the LP dispersion curve transitions from exciton-like to photon-like. Consequently, polariton-phonon scattering becomes less efficient due to a reduced scattering rate at the bottleneck and the LP lifetime also decreases. Polariton relaxation to the ground state is thereby inhibited. One approach to solving this problem is by introducing polariton-electron scattering, which has been shown both theoretically [15] and experimentally [16-17] to be more efficient than polariton-phonon scattering. Experiments involving polariton-electron scattering typically involve the introduction of a suitable of 2-dimensional electron gas (2DEG) within a wide QW through photo-excitation in an adjacent narrow QW [16-17]. In an electrically injected device with a QW active region, direct doping of the wells would result in broadening of the emission linewidth due to ionized impurity scattering. A more elegant approach would be to introduce the electrons into the QWs through modulation doping in the barrier regions adjacent to the QWs. In what follows, we report on an experimental study of the effects of polariton-electron scattering on polariton relaxation in a modulation-doped multi-QW $In_{0.1}Ga_{0.9}As$/GaAs microcavity diode operating in the strong coupling regime. It is observed that the relaxation bottleneck is suppressed, and the rate of polariton relaxation to the ground state is enhanced.

The microcavity device used in this study, shown in Fig. 1, was grown by molecular beam epitaxy on a n-doped GaAs (001) substrate and consists of an undoped (n ~ $10^{15}$ cm$^{-3}$) 5λ/2 GaAs cavity surrounded by 32 periods of Si-doped (n = 2 x $10^{18}$ cm$^{-3}$) GaAs / $Al_{0.85}Ga_{0.15}As$ forming the bottom distributed Bragg reflector (DBR) and 25 periods of Be-doped (p = 4 x $10^{18}$ cm$^{-3}$) GaAs / $Al_{0.85}Ga_{0.15}As$ forming the top DBR. Four periods of the DBR adjoining the cavity on both sides were left undoped to minimize doping related losses. Four pairs of undoped 10 nm $In_{0.1}Ga_{0.9}As$/10 nm GaAs QWs are grown at the antinodes of the cavity photon field to maximize the interaction between the QW excitons and the cavity photons. A 1 nm thick region at a distance of 10 nm below each pair of QWs was Si-doped n-type. Three samples were investigated with a Si-doped concentration of n = 0, 1x$10^{17}$ cm$^{-3}$ and 2x$10^{17}$ cm$^{-3}$ respectively. Since at these levels the modulation doping is highly degenerate, the impurity band overlaps with the conduction band and the dopants require no energy for ionization and thus the corresponding thermally ionized electron sheet densities are $n_e$ = 0, 1x$10^{10}$ and 2x$10^{10}$ cm$^{-2}$ for the 3 samples. Standard photolithography and wet chemical etching techniques were employed to form a mesa of 50 μm diameter. Ring-shaped annular Pd/Zn/Pd/Au p-contact and Ni/Ge/Au/Ti/Au n-contact were formed, for hole and electron injection, respectively, to complete device fabrication. The contacts were deposited on the first doped Bragg pair to minimize Joule heating. The top p-contact has an annular width of 10 μm and therefore covers more than half of the device mesa area and ensures a nearly uniform injection.

The dispersion of the resonant modes was measured at 30K by angle resolved electroluminescence, with an angular resolution of 1°, at an injection current density of 0.8 A/cm$^2$ with a high resolution monochromator (spectral resolution ~ 0.03 nm) and detected with a photomultiplier using phase sensitive lock-in amplification. The measured spectra for the sample with a modulation doped sheet electron density of $n_e$ = 1x10$^{10}$ cm$^{-2}$ are shown in Fig 2(a). The result shows two distinct polariton peaks, with the LP peak asymptotically approaching the exciton energy at larger angles. Signature of a weak middle polariton (MP) branch is also observed, which has been attributed to the effect of charged excitons or trions (X$^-$) [16,18]. The resonances obtained from the electroluminescence spectra are plotted in Fig. 2(b) together with the calculated dispersion curves (solid lines) of the upper polariton (UP), MP and lower polariton (LP) branches by using the one-to-one correspondence between the emission angle of the out-coupled photons and the in-plane wave number of the polaritons. The interaction between the cavity and exciton modes was modeled with a coupled harmonic oscillator Hamiltonian:

$$\begin{pmatrix} E_{cav} - j\Gamma_{cav} & V_{X_{hh}} & V_{X^-} \\ V_{X_{hh}} & E_{X_{hh}} - j\Gamma_{X_{hh}} & 0 \\ V_{X^-} & 0 & E_{X^-} \end{pmatrix} \qquad (1)$$

where E$_{cav}$ is the bare cavity photon resonance and $V_{X_{hh}}$ and $V_{X^-}$ are the interaction potentials of the heavy-hole exciton and trion with the cavity mode. The measured values of heavy-hole exciton energy (E$_{Xhh}$ =1.3779 eV), exciton linewidth (Γ$_{Xhh}$= 1.2 meV) and cavity photon linewidth (Γ$_{cav}$ =0.21 nm) have been used in the analysis. Only the interaction between the cavity photon and the heavy-hole exciton is considered because the light-hole exciton transition is energetically situated 15 meV above the hh-exciton transition due to biaxial strain in the quantum wells. The measured results exhibit excellent agreement with the values obtained from the calculations for interaction potentials $V_{X_{hh}}$ = 5.1 meV and $V_{X^-}$ = 1.8 meV and a cavity to heavy-hole exciton detuning δ of -3.5 meV.

In order to determine the effect of the polariton-electron scattering in suppressing the bottleneck and enhancing the rate of polariton relaxation to the $k_\parallel \sim 0$ states, we have calculated the polariton occupancy in k-space as a function of $n_e$ from the measured angle-resolved electroluminescence data at an injection current density of 4A/cm$^2$ by using the formula, $I_{LP}(k_\parallel) = \eta N_{LP}(k_\parallel)|C(k_\parallel)|^2 M/\tau_C$ [19], where η is the collection efficiency, $\tau_C/|C(k_\parallel)|^2$ is the radiative lifetime of the LPs, M is the number of transverse states included in the detection cone and $|C(k_\parallel)|^2$ is the photon fraction at a wave-vector $k_\parallel$. The number of states within the detection cone is given by[7] $M = \frac{D^2}{16}(k_0\Delta\theta)^2$, where D is the diameter of the emission spot, $k_0 = 2\pi/\lambda$ and Δθ is the detection half angle. $\tau_C$ is estimated from the cavity Q, and the value of η is estimated by replacing the sample with a source of known power (Ti:sapphire laser with suitable attenuation). The LP number density per k-state is plotted in Fig. 3(a). A pronounced relaxation bottleneck at $k_\parallel \sim 1.17 \times 10^4$ cm$^{-1}$ is observed for the sample with $n_e = 0$. However for samples with $n_e = 1 \times 10^{10}$ and $2 \times 10^{10}$ cm$^{-2}$, the population of the ground state increases, while the reservoir population remains almost unchanged, and the bottleneck is removed.

To investigate the dependence of occupation of different k-states on injected current density and the possibility of polariton condensation, we measured the angle-resolved electroluminescence at 30K for the sample with $n_e = 1 \times 10^{10}$ cm$^{-2}$. The data are shown in Fig. 3(b). At low injection densities, as mentioned earlier, a distribution without any bottleneck is observed with comparable occupancies between $k_\parallel = 0$ and $1.23 \times 10^4$ cm$^{-1}$. At higher injection currents, the occupancy of the ground state shows a super-linear increase, while that of the reservoir shows very little increase. Although a bimodal distribution (with a massively occupied ground state) associated with polariton lasing is not observed, the occupancy profile can be characterized by the Maxwell-Boltzmann distribution: $N = N_0 \exp(-(E(k)-E(0))/kT)$, (where $E(k)$ is the LP energy at $k_\parallel$ and $N_0$ is the occupancy at $k_\parallel = 0$). It is evident that polariton-electron scattering plays a role similar to polariton-phonon scattering and thermalizes the LP distribution [15].

Finally, the dependence of light output on the injection current density was investigated by measuring the electroluminescence at $k_\parallel \sim 0$ as a function of injection current (pulsed to avoid device heating). With increasing current density the UP and LP polariton peaks progressively come closer in energy (Fig 4(a)) and merge into a single emission line, signaling the transition to the weak coupling regime. The two peak energies merge at the cavity photon energy of 1.374 eV at θ=0° (see dispersion curve in Fig.2(b)) and for an injection current density of 22 A/cm$^2$. At this point the normal modes of the system become cavity photons and QW excitons and the corresponding carrier density is $4.2 \times 10^{10}$ cm$^{-2}$, in agreement with previous reports [20]. As the current density increases, linewidth narrowing of the LP emission is not observed. The integrated electroluminescence from the lower polaritons was measured as a function of injection current density and the data are shown in Fig.4(b). In the absence of polariton-electron scattering ($n_e = 0$), the intensity grows linearly with injection, which is a signature of polariton-phonon scattering. In contrast, in the presence of such scattering, the measured intensity grows super-linearly with increasing current density (in agreement with the increase in polariton occupation at $k_\parallel \sim 0$). A quadratic increase in the LP emission intensity is observed in a small range of current density (15 A/cm$^2$ ≤ J ≤ 22 A/cm$^2$). The observed super-linear dependence on current is attributed to enhanced relaxation of LPs due to polariton-electron scattering, in agreement with the data of Fig.

3(a). The quadratic dependence at the highest injection current densities suggests that the relaxation in this regime is dominated by polariton-polariton scattering [17].

In conclusion, we have investigated the effect of polariton-electron scattering on the polariton relaxation dynamics and output characteristics of an electrically injected GaAs-based microcavity p-i(MQW)-n diode in the strong coupling regime. Excess electrons were provided in the quantum wells by modulation doping with Si. It is observed that the relaxation bottleneck is suppressed and the emission develops a superlinear dependence on injection current density with increasing density of electrons available for polariton-electron scattering. Polariton lasing is not observed, which suggests that polariton-electron scattering is not adequate for attaining ground state quantum degeneracy.

The work is supported by the Air Force Office of Scientific Research under Grant FA9550-09-1-0634.

**Figure Captions:**

*Figure 1* (color online) Schematic representation of fabricated GaAs-based modulation doped quantum well-microcavity diode with 4 pairs of $In_{0.1}Ga_{0.9}As$ / GaAs quantum wells placed at the antinodes of the $5\lambda/2$ cavity region and 4 δ-doped layers. The quantum wells are modulation doped with electrons from Si dopants in the barrier regions.

*Figure 2* (color online) Strong coupling effects measured with a 50 μm diameter device with sheet electron concentration of $n_e = 1\times10^{10}$ cm$^{-2}$: (a) angle-resolved electroluminescence at 30K showing the upper, middle and lower polariton resonance peaks; (b) polariton dispersion curves obtained from (a). The solid curves represent the polariton dispersions calculated from a coupled harmonic oscillator model (see text).

*Figure 3* (color online) (a) Occupancy of lower polariton branch (LPB) as a function of in-plane wave vector for different modulation doping density ($n_e = 0$ and $1\times10^{10}$ cm$^{-2}$) obtained from angle-resolved electroluminescence; (b) occupancy of LPB for various injection currents at T = 30K in a device with doping density $n_e = 1\times10^{10}$ cm$^{-2}$.

*Figure 4* (color online) (a) LP and UP emission peak energies (left) and EL spectra (right) as a function of injection current density; (b) integrated emission intensity measured in the normal direction versus injection current density for $n_e = 0, 1\times10^{10}$ cm$^{-2}$ and $2\times10^{10}$ cm$^{-2}$.

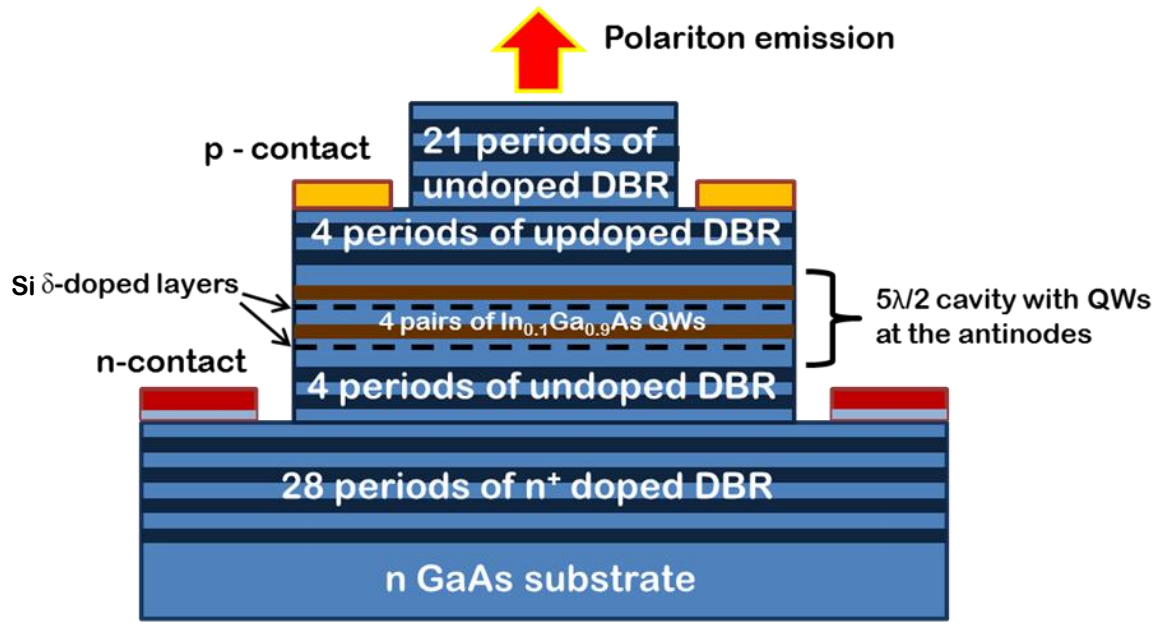

Schematic of the device

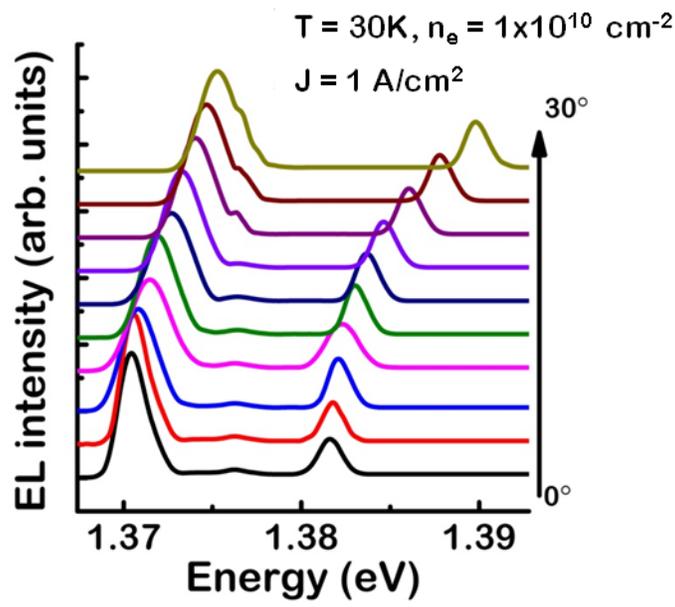

(a)

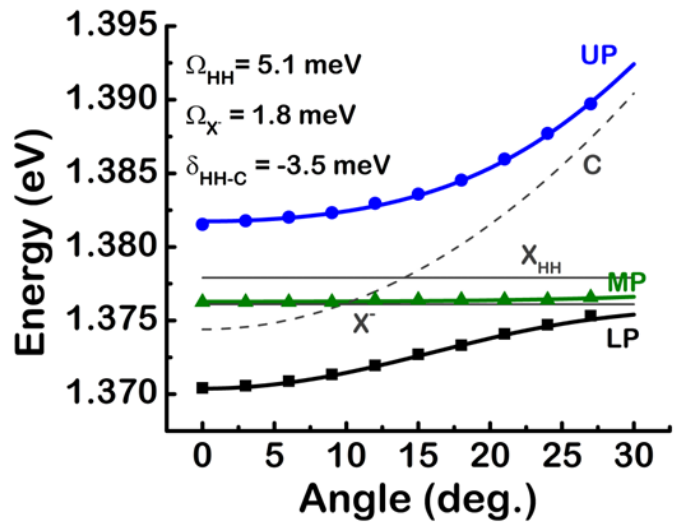

(b)

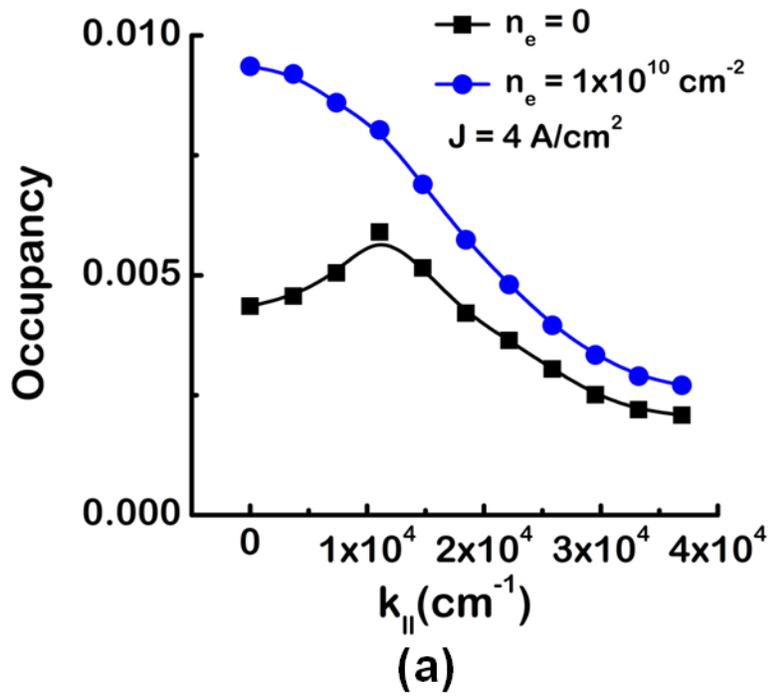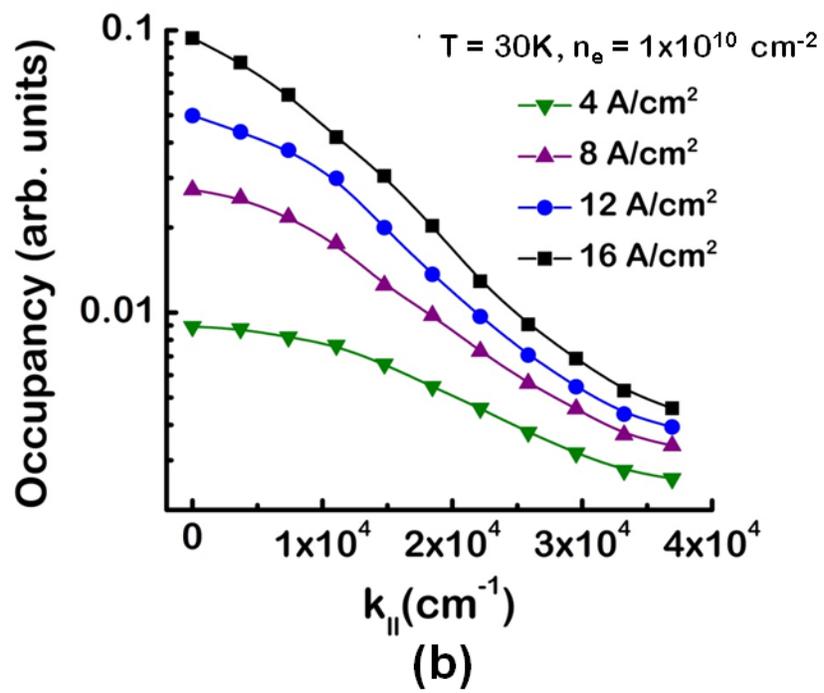

Das *et al*, Fig. 3 of 4

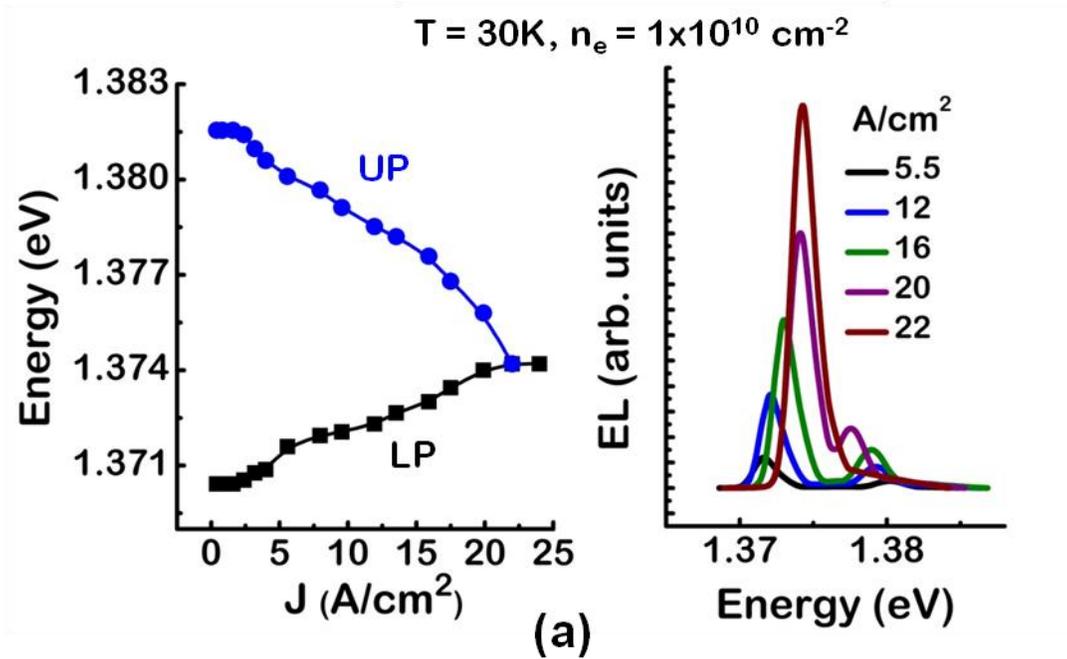

(a)

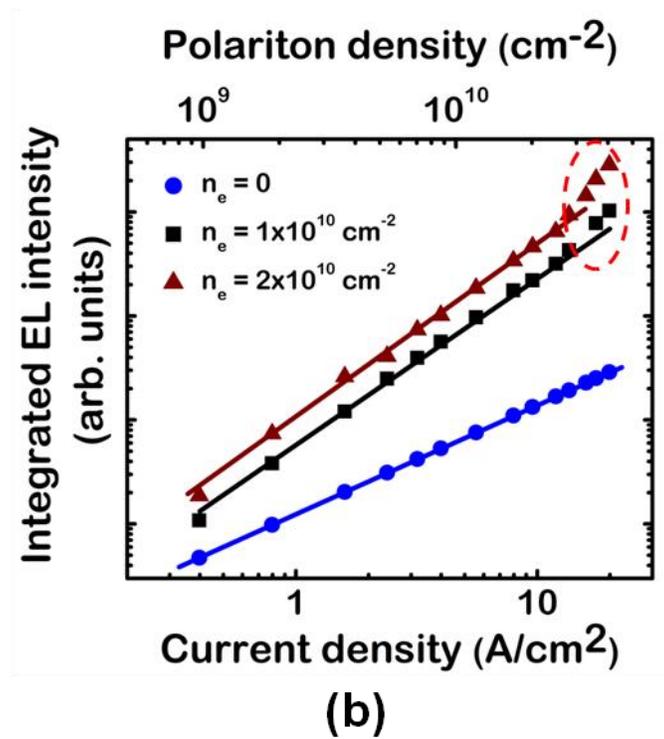

(b)

Das *et al*, Fig. 4 of 4